\pdfoutput=1
\documentclass[twocolumn,showpacs,floatfix,prb,superscriptaddress]{revtex4}

\usepackage[final]{graphicx}
\DeclareGraphicsExtensions{.pdf}
\usepackage{amsmath,amssymb,bbold,bm}
\usepackage{float}
\usepackage{hyperref}

\newcommand{\bk}{{\bf k}}

\newcommand{\bq}{{\bf q}}

\newcommand{\br}{{\bf r}}
\newcommand{\bR}{{\bf R}}

\newcommand{\bK}{{\bf K}}
\newcommand{\Tr}{{\rm Tr}}

\def\id{\mathbb{1}}

\begin{document}

\title{Quasiparticle spectroscopy as a probe of the topological phase in graphene with heavy adatoms}

\author{Paul Soul\'{e}}
\affiliation{Laboratoire de Physique Th\'{e}orique et Mod\`{e}les statistiques, Universit\'{e} Paris-Sud, 91405 Orsay, France}
\affiliation{Department of Physics and Astronomy, University of
British Columbia,Vancouver, BC, Canada V6T 1Z1}
\author{M. Franz}
\affiliation{Department of Physics and Astronomy, University of
British Columbia,Vancouver, BC, Canada V6T 1Z1}
\affiliation{Quantum Matter Institute, University of British Columbia, Vancouver BC, Canada V6T 1Z4}

\begin{abstract}
Electrons in graphene with heavy adatoms (such as In or Tl) have been predicted to form a 2D topological insulator phase with a substantial spectral gap potentially suitable for future practical applications. In order to facilitate the ongoing experimental efforts to identify this phase we perform a theoretical study of its spectral properties in a model graphene system with randomly distributed adatoms. Our extensive modeling shows that random heavy adatoms produce a full spectral gap (as opposed to a mobility gap) accompanied by distinctive quasiparticle interference patterns observable by means of Fourier-transform scanning tunneling spectroscopy. 

\end{abstract}

\maketitle

Despite their pivotal role in the ``topological revolution'' that transpired  in condensed mater physics  in recent years \cite{moore_rev,hasan_rev,qi_rev,franz_book} 2D topological insulators (TIs) have thus far largely failed to deliver on their promise to become  a testbed for fundamental new concepts and a platform for exciting practical applications. The reason behind this lies in the lack of widely available 2D TI materials. The existing known 2D TIs include HgTe/CdTe quantum wells \cite{konig} and InAs/GaSb quantum wells \cite{knez}, which however require specialized fabrication techniques and have not, thus far, caught on as convenient and widely available platforms for broad experimentation. This is in contrast to 3D TIs \cite{franz_book}  where dozens of confirmed materials exist and the prototype Bi$_2$Se$_3$, Bi$_2$Te$_3$ materials are straightforward to grow and widely available.

The historically first and conceptually simplest 2D TI system is based on the Kane-Mele model \cite{kane1} for graphene with spin-orbit coupling (SOC). Although the  intrinsic SOC strength is too small to bring about this phase in pristine graphene it has been suggested that the effect can be amplified manyfold by depositing a dilute concentration of certain heavy adatoms. Specifically, graphene with a modest  $\sim 6$\% concentration of In and Tl adatoms is predicted to form a TI with an estimated gap of 7 and 21 meV, respectively.\cite{weeks1} These adatoms' outer electrons are in $p$ shells and in essence act as local sources of strong SOC for low-energy Dirac electrons in graphene. Potentially much larger gaps can be achieved by using transition metal elements with active $d$ orbitals such as Ir and Os, although the detailed microscopic mechanism is somewhat different here.\cite{hu1}

Although conceptually simple and straightforward to implement, the proposal to generate a 2D TI from graphene with adatoms has not yet been experimentally realized. Transport experiments\cite{folk1} on graphene flakes with very small concentrations of In ($<0.02$\%) have confirmed the predicted doping dependence (each In adatom donates $\sim 1$ electron) but were unable to confirm the transition into the topological state which one expects only at higher adatom densities. Preliminary scanning tunneling microscopy\cite{burke1} (STM) studies of Tl on graphene grown on SiC substrate indicated the `hollow' adsorption site (in the middle of the hexagonal plaquette) as predicted but could not resolve the spectral gap characteristic of a 2D TI. Angle resolved photoemission\cite{dama1} (ARPES) on similar samples observed the effect of doping as well as increased line broadening, but again failed to discern any clear signature of an excitation gap.  

In order to assist the ongoing experimental efforts aimed at identifying the 2D topological phase in graphene with adatoms we undertake here a program of theoretical modeling of its spectral properties in the experimentally relevant regime of {\em randomly distributed} adatoms. Aside from detailed predictions that we develop for STM and ARPES our study yields two important qualitative insights. First, we find that SOC generated by randomly distributed heavy adatoms produces a {\em full spectral gap} (as opposed to a mobility gap). This feature was not apparent from the original transport calculations in the disordered regime\cite{weeks1,hu1,shevtsov1}  although more recent work\cite{niu1} indicated that this might be the case. Second, we identify unique signatures of the SOC observable by Fourier-transform scanning tunneling spectroscopy (FT-STS). These take the form of quasiparticle scattering patterns that are prohibited by symmetries in graphene with ordinary potential scatterers.\cite{shon} Our results 
thus 
identify 
ARPES in 
combination with FT-STS as ideal tools for observing the topological phase in graphene with heavy adatoms.

In this study we focus on the simpler and physically more transparent model appropriate for In and Tl adatoms\cite{weeks1} defined by the lattice Hamiltonian $H=H_t+\sum_I\delta H_{I}$
with
\begin{eqnarray}  \label{h1}
  H_t &=& -t \sum_{\langle {\bf r r'}\rangle}(c^\dagger_{{\bf r}}c_{{\bf r'}} + {\rm h.c.}) +\sum_\br w_\br c^\dagger_{{\bf r}}c_{{\bf r}},
\\
\delta H_I&=&-\delta \mu\sum_{{\bf r}\in I} c^\dagger_{{\bf r}}c_{{\bf r}}
+\lambda_{\rm so} \sum_{\langle\langle {\bf r r'}\rangle\rangle\in I}(i \nu_{\bf r r'}c^\dagger_{{\bf r}}s^z c_{{\bf r'}} + {\rm h.c.}).
 \nonumber
\end{eqnarray} 
Here $H_t$ describes the usual nearest-neighbor electron hopping on the graphene honeycomb lattice with $t\simeq 2.7$eV and $w_\br$ denoting weak random disorder (unrelated to adatoms) coming from the substrate or other sources. $\br$ denotes the lattice site and the electron spin is treated implicitly (i.e.\ we view $c_\br$ as a two-component spinor, $s^z$ is the Pauli matrix). In the second line $I$ labels the random plaquettes occupied by adatoms. The first term in $\delta H_I$ describes the chemical potential that screens charge from the adatoms, while the second term captures the local intrinsic  spin-orbit coupling induced by electrons hopping from graphene to an adatom and back. We neglect the Rashba coupling, which has been shown unimportant.\cite{weeks1} In addition, $\nu_{\br\br'}=+1$ for hops clockwise around the plaquette and $-1$ counterclockwise. Realistic parameters for Tl adatoms are $\lambda_{\rm so}=0.02t$ and $\delta\mu=0.1t$.

According to the previous work\cite{weeks1,niu1} we expect the SOC induced by In and Tl adatoms to open a gap in the electron excitation spectrum at the Dirac point. It is thus useful to start our discussion by considering the effective low-energy theory obtained by projecting Hamiltonian (\ref{h1}) to the vicinity of the two Dirac momenta $\pm\bK=\pm(4\pi/3\sqrt{3}a,0)$ with $a$ the separation between nearest carbon atoms. The low-energy Hamiltonian reads $H^{\rm eff}=\int d^2r \psi^\dagger(\br)(h_0+h')\psi(\br)$ with 
\begin{eqnarray}
  h_0 &=& -i\hbar v\left(
\tau^z\sigma^x\partial_x+\sigma^y\partial_y\right),
  \label{h0}\\
h'&=&\sum_j\left(-3\delta\mu+\Lambda_{\rm so}\tau^z\sigma^z s^z\right)S_0\delta(\br-\bR_j).
 \nonumber
\end{eqnarray} 
Here ${\bm\tau}$ and ${\bm\sigma}$ are Pauli matrices acting in the valley and sublattice space, respectively, $v$ represents the Fermi velocity, $\Lambda_{\rm so}=3 \sqrt{3} \lambda_{\rm so}$, $S_0$ is the area of the unit cell and $\bR_j$ denotes the random adatom positions. The 8-component spinor $\psi(\br)$ describes the low-energy electron field in combined valley, sublattice and spin space. For simplicity we neglect the substrate disorder here but we come back to it later. Upon Fourier transforming the Hamiltonian takes the standard form of a disorder problem,\cite{doniach1}
\begin{equation}\label{heff}
H^{\rm eff}=\sum_\bk\psi^\dagger_\bk h_\bk\psi_\bk
+\sum_{\bk\bq}\psi^\dagger_{\bk+\bq}\rho_\bq U_\bq\psi_\bk,
\end{equation}
with $h_\bk=v(\tau^z\sigma^x k_x+\sigma^y k_y)$, $\rho_\bq=\sum_je^{-i\bR_j\cdot\bq}$ and  $U_\bq=(-3\delta\mu+\Lambda_{\rm so}\tau^z\sigma^z s^z)S_0/S$ and $S$ the area of the system.

We are interested in the disorder-averaged electron propagator 
\begin{equation}\label{geff}
g(\bk,\omega)=\left[g_0(\bk,\omega)^{-1}-\Sigma(\bk,\omega)\right]^{-1}
\end{equation}
where $g_0(\bk,\omega)=(\omega+i\delta-h_\bk)^{-1}$ is the propagator of the clean system with $\delta=0^+$ while $\Sigma(\bk,\omega)$ represents the disorder self energy.
For weak disorder we can evaluate the latter using the standard Born series, which corresponds to the expansion in powers of $U_\bq$. To first order we obtain simply\cite{doniach1}  
\begin{equation}\label{sig1}
\Sigma^{(1)}(\bk,\omega)=N_IU_{\bq=0}=n_I(-3\delta\mu+\Lambda_{\rm so}\tau^z\sigma^z s^z),
\end{equation}
where $N_I$ is the total number of impurities (adatoms) and $n_I=N_I(S_0/S)$ is their number density. The key point to notice here is that while the scalar term $-3\delta\mu$ in $\Sigma^{(1)}$ merely shifts the overall chemical potential the SOC term opens up a spectral gap at the Dirac point with the amplitude $\Delta_{\rm so}=n_I\Lambda_{\rm so}$. Therefore, the first order Born correction, which is often neglected as unimportant for scalar disorder potential, leads to an important qualitative change in the spectral properties of the system. Furthermore, to this order the effective disorder averaged Hamiltonian $h_\bk^{(1)}=h_\bk+\Sigma^{(1)}(\bk,0)$ is identical to the Kane-Mele model \cite{kane1} and describes a $Z_2$ topological insulator with bulk gap and protected gapless edge states.

The second order Born expansion gives
\begin{align}\label{sig2}
\Sigma&^{(2)}(\bk,\omega)=N_I\sum_\bq U_{\bk-\bq}g_0(\bq,\omega)U_{\bq-\bk}\\
&=n_I\left({3\delta\mu-\Lambda_{\rm so}\tau^z\sigma^z s^z\over \Lambda}\right)^2{\omega\over 4\pi}\left\{\ln{\omega^2\over\Lambda^2}-i\pi\rm{sgn}(\omega)\right\}, \nonumber
\end{align}
where $\Lambda=v/\sqrt{S_0}\simeq t$ is the high-energy cutoff for Dirac fermions. For the relevant frequencies $\omega\simeq \Delta_{\rm so}$ we observe that $\Sigma^{(2)}$ represents a small correction to $\Sigma^{(1)}$ as long as $n_I(\delta\mu/\Lambda)^2,n_I(\Lambda_{\rm so}/\Lambda)^2\ll 1$, which we expect to be always true. Higher terms in the Born expansion will be down by additional powers of these small parameters and can therefore be neglected. We conclude on this basis that random distribution of heavy adatoms will indeed open a gap $\Delta_{\rm so}\simeq n_I\Lambda_{\rm so}$ in the spectrum of Dirac fermions.  In addition, the disorder induces quasiparticle lifetime broadening $\Gamma={\rm Im}\Sigma$ already apparent  in Eq.\ (\ref{sig2}). We expect the disordered system to remain in the topological phase as long as $\Gamma\lesssim\Delta_{\rm so}$ and the chemical potential stays inside the gap.   

The gap predicted to exist in graphene with randomly distributed heavy adatoms should be directly observable by various spectroscopies such as ARPES and STS and in transport measurements. Such observation alone would provide a strong support for the notion of the topological phase but would not constitute a definitive proof. Detection of quantized edge transport would provide definitive evidence but is complicated by the need to position the chemical inside the gap. As a plausible alternative to transport measurements we study here quasiparticle interference patterns, observable by FT-STS, which we show contain unique signatures of the  SOC origin of the spectral gap. 

An FT-STS experiment\cite{crommie1,davis1} probes the local density of states, $n(\br,\omega)$, at a large number of real-space locations $\br$ on the sample surface. The spatial Fourier transform of this signal $n(\bq,\omega)$, referred to as FT-LDOS, can be related to the full electron propagator $G(\br,\br';\omega)$ as
\begin{equation}\label{n1}
n(\bq,\omega)=-{1\over \pi}\Im\int d^2r e^{-i\br\cdot\bq}\Tr[G(\br,\br;\omega)].
\end{equation}
Here the trace is taken over spin and orbital quantum numbers and  $\Im$ denotes the strength of the branch cut across the real frequency axis $\Im f(\omega)\equiv [f(\omega+i\delta)-f(\omega-i\delta)]/2i$. In the limit of weak random potential, the interesting $\bq$-dependent part of the FT-STS signal can be expressed in a simple factorized form,\cite{capriotti1}
\begin{eqnarray}
\delta n(\bq,\omega) &=&-{1\over \pi} \rho_\bq  
{\Im} [\Lambda(\bq,\omega)],\label{born1} \\
\Lambda(\bq,\omega) &=&\sum_\bk\Tr[ G_0(\bk,\omega)U_\bq G_0(\bk-\bq,\omega)],
\label{lam1}
\end{eqnarray}
where $G_0(\bk,\omega)$ is the electron propagator in the absence of disorder. Since $\rho_\bq$ is the
Fourier transform of a random potential one expects it to be a
featureless function of $\bq$. $\Lambda(\bq,\omega)$, on
the other hand, represents the response of the underlying {\em clean}
system and contains, in general, prominent features as a function of
$\bq$ that can be used to study its properties.

Compared to the standard theoretical treatment\cite{capriotti1} of FT-LDOS where disorder can be neatly separated from the underlying `clean' system our problem exhibits
a slight difficulty in that adatoms provide both the source of disorder and of the spectral gap that we would like to probe. To address this complication we follow a two-pronged strategy. First, we use an analytical approach in which we focus on the low-energy theory (\ref{h0}) and take the first-order disorder-averaged Hamiltonian $h_\bk^{(1)}$ to describe the underlying clean system. We then assume that residual disorder, not contained in the first-order Born approximation, plus any disorder not related to adatoms (e.g.\ substrate) is sufficiently weak and permits the use of Eq.\ (\ref{born1}) to calculate the interference pattern. Second, to confirm the validity of this approximate analytical treatment, we consider the full lattice Hamiltonian (\ref{h1}) with realistic parameters. We perform exact numerical diagonalizations on finite clusters for specific random adatom configurations and compute FT-STS response with no approximations directly from Eq.\ (\ref{n1}).

\begin{figure}[t]
 \centering
 \includegraphics[scale=0.8,keepaspectratio=true]{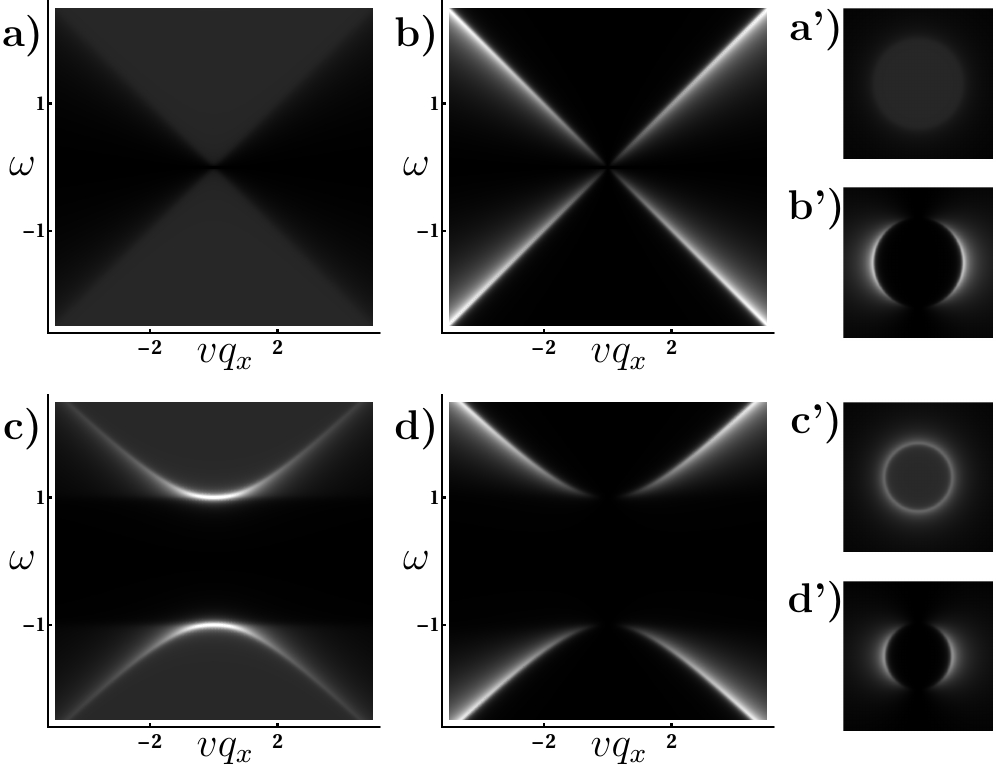}
 \caption{Grayscale  plots of $|\Im \Lambda^{\tau\tau'}(\bq,\omega)|$ from Eq.\ (\ref{Intra2}) and (\ref{Inter2}). Top panels a) and b) show the intravalley and intervalley $\omega$-$q_x$ maps for pristine graphene ($\Delta_{\rm so}=0$). Bottom panels c) and d) represent the same maps for $\Delta_{\rm so}=1$. Panels a'), b'), c'), and d') display transverse sections in the $q_x$-$q_y$ plane of the corresponding plots for $\omega=1.5$. We use one grayscale for all intravalley features, and an other one for intervalley plots.}
\label{fig1}
\end{figure}
The analytical approach consists of evaluating the momentum sum in Eq.\ (\ref{lam1}) with $G_0(\bk,\omega)=[\omega+i\delta-h_\bk^{(1)}]^{-1}$ in the low energy approximation. If $|q| \ll a^{-1}$ only scattering within the same valley contributes to the sum, whereas scattering from one valley to another appears for $\bq$ close to the corners of the Brillouin zone. To calculate $\Lambda^{\tau\tau'}(\bq,\omega)$ we use the unperturbed one-particle Green's function $G_0(\bk,\omega)=(\omega+i\delta-h_\bk-\Delta_{\rm so}\tau \sigma^z s^z)^{-1}$ where we have subsumed the shift $-n_I3\delta\mu$ into the bulk chemical potential and $\tau=\pm 1$ is the valley index. We assume here for simplicity that the disorder potential is non-magnetic and slowly varying on the lattice spacing scale such that $U_\bq=u_0\id$ in Eq.\ (\ref{lam1}).

For the intravalley term, switching to Matsubara frequencies $i \omega_n$, we obtain
\begin{gather}\label{Intra1}
\Lambda^{++}(\bq,i\omega_n)= 8 \sum \limits_k \frac{(i \omega_n)^2+\Delta_{\rm so}^2+ v^2 \bk (\bk - \bq) }{D}, \\
\nonumber
 D = \left( \omega_n^2 + \Delta_{\rm so}^2 + v^2 \bk^2 \right) \left( \omega_n^2 + \Delta_{\rm so}^2 + v^2 (\bk-\bq)^2 \right). 
\end{gather}
Integrals of this type can be computed in a similar way as for pristine graphene\cite{tami3} by means of Feynman parametrization\cite{peskin,tami1}. We find 
\begin{equation}\label{Intra2}
\Lambda^{++}= \frac{2S}{\pi v^2} \left[ \ln (\frac{\Lambda^2}{\omega_n^2+\Delta_{\rm so}^2} ) + 2 g(z) - \frac{8 \Delta_{\rm so}^2}{v^2 q^2}f(z) \right], 
\end{equation}
where  $z=4 [(i \omega_n)^2 - \Delta_{\rm so}^2]/{v^2 q^2}$ and we defined functions $f(z)=\frac{1}{\sqrt{z-1}}\arctan \bigl( \frac{1}{\sqrt{z-1}} \bigr) $ and $g(z)=(z-1)f(z)$. We emphasize that $f(z)$ has a singularity at $z=1$ whereas $g(z)$ does not.
Therefore, in the absence of SOC, the FT-LDOS has no singularities in the intravalley response\cite{tami3,bena1}. However, the term proportional to $\Delta_{\rm so}^2$ is singular when $z=1$ or equivalently when $\varepsilon(q/2)=\omega$, where $\varepsilon(k)=\pm\sqrt{v^2k^2+\Delta_{\rm so}^2}$ is the dispersion relation of $h^{(1)}$. Those singularities arise from elastic backscattering terms in the sum\ (\ref{lam1}) when $\bq=2\bk$ and $\omega=\varepsilon(\bk)=\varepsilon(\bk-\bq)$. Pseudospin chirality conservation prohibits this intravalley backscattering in pristine graphene because incoming and outgoing quasiparticles have an opposite pseudospin direction.\cite{shon,mallet1} In the presence of the SOC mass term $\Lambda_{\rm so}$, however, the chirality conservation is broken and intravalley backscattering close to the gapped region is allowed. 

The intervalley component for $|\bq-\bK| \ll a^{-1}$ is obtained from Eq.\ (\ref{lam1}) using $\tau=+1$ and $\tau'=-1$ for the left and right $G_0$ term, respectively. We find 
\begin{equation}\label{Inter2}
 \Lambda^{+-}(\bq,i\omega_n)= \frac{S}{\pi v^2} \left[2 \frac{q_x^2}{q^2}\left(1-zf(z)\right) -1 \right].
\end{equation}
Here, the surface $z=1$ is singular even without SOC, but the amplitude is angle dependent. Singularities arise from scattering of quasiparticles from one valley to the other, but here the overlap of incoming and outgoing quasiparticles' pseudospins depends on $\mathbf q$ direction.

In Fig.\ \ref{fig1} we plot the FT-STS signal $|\Im[\Lambda^{\tau\tau'}]|$ based on Eqs.\ (\ref{Intra2}) and (\ref{Inter2}). Without SOC, the intravalley signal is non-singular and  barely visible whereas a linear dispersion with slope $v/2$ appears in the intervalley signal. When the SOC is present, we see a qualitative change in the maps. Now a parabolic dispersion is clearly visible both in the intra- and the intervalley FT-LDOS with a gap $2\Delta_{\rm so}$ separating the two bands. We have also computed numerically $\Lambda(\bq,i \omega)$ from Eq.\ (\ref{lam1}) away from the low energy approximation of $h_0$, and checked that the characteristic features described above remain unchanged as long as $\Delta_{\rm so} \lesssim t$. Finally, a rapidly oscillating disorder potential might have different amplitudes on $A$ and $B$ sublattices such that $U_\bq = u_0\id+ \alpha_\bq \sigma^z$ in Eq.\ (\ref{lam1}). One can check, however, that the $\sigma^z$ term does not contribute to the intravalley response and 
affects only the 
amplitude of the singularities in the intervalley term, in such a way that our above statements remain true.

\begin{figure}[t]  
 \includegraphics[scale=0.85,keepaspectratio=true]{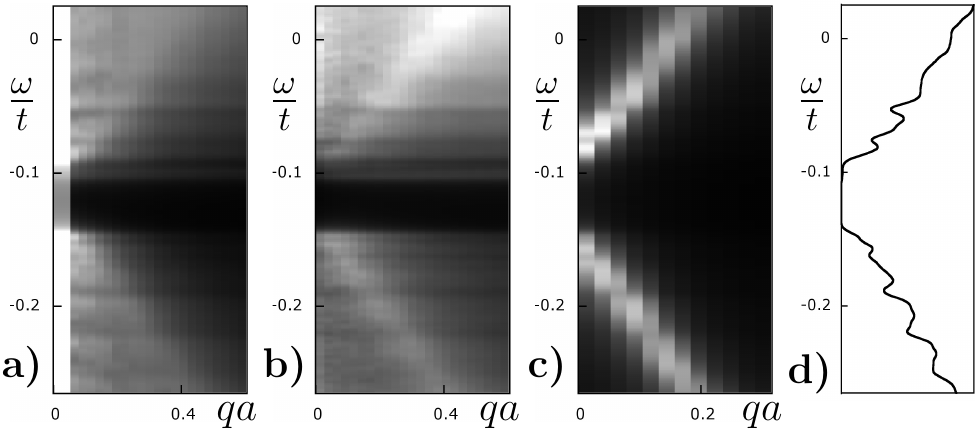}
 \caption{Numerical computation of $|n(\bq,\omega)|$ for the lattice model of Eq.\ (\ref{h1}) on a $80\times80$ periodic cluster with $\lambda_{\rm so}=0.04t$ and $\delta\mu=0.2t$. Panel a) and  b) show closeups of the intravalley and  intervalley FT-LDOS, respectively.  Panel c) represents the spectral function $A(\bq,\omega)$ and d) the total density of states $n(\omega)$.}
\label{fig2}
\end{figure}
Even though our computations above were performed for an averaged adatom distribution, we believe that the characteristic signal of the topological phase can be observed in FT-STS experiments. In order to support this claim, we carried out exact numerical simulations based on the lattice model of Eq.\ (\ref{h1}) for specific disorder configurations. These computations have the advantage of not relying on the weak disorder or low energy approximations. In addition, no average over disorder configurations is performed before computing the FT-STS signal, just like in real experiments. The FT-LDOS is evaluated from Eq.\ (\ref{n1}) which can be manipulated into the more convenient expression
\begin{equation}
 n(\br,\omega)=-{1\over \pi}\Im \sum \limits_i  \frac{|\Psi_i(\br)|^2}{\omega+i\delta-E_i}.
\end{equation}
where $\Psi_i(\br)$ and $E_i$ are the eigenvectors and eigenvalues of our one-body Hamiltonian (\ref{h1}) computed by means of exact numerical diagonalization. In Fig.\ \ref{fig2}, we present our results computed on clusters of $80\times80$ unit cells for parameters $\lambda_{\rm so}=0.04t$ and $\delta\mu=0.2t$, close to realistic values. We consider here an adatom coverage of $n_I=0.2$ and an uncorellated random potential $w_r\in [-0.04t, 0.04t]$. The latter has in fact little effect because it remains much smaller than the disorder induced by adatoms whose variance is about $3 n_I \delta\mu^2$. In order to achieve better resolution, we show the FT-LDOS and the spectral function signal as angular averages over circular regions around the $\bq=0$ or $\bq=\bK$ points. In addition, we average each quantity over 10 independent realization of disorder. 

Fig.\ \ref{fig2}a,b shows  a clear energy gap in the intra- and intervalley components of  FT-LDOS. This gap is somewhat smaller than $2\Delta_{\rm so}=6\sqrt{3} n_I \lambda_{\rm so} \approx 0.083 t$ obtained in our approximate analytical calculation, but remains open for each of our ten disorder configurations. Moreover, this gap also appears in the spectral function and in the total density of states. This indicates that a random distribution of adatoms on the graphene sheet not only opens a mobility gap as demonstrated  by Weeks and coworkers\cite{weeks1}, but produces a full spectral gap observable through ARPES and FT-STS experiments. One can also perceive the parabolic electron dispersion in the most intense regions of FT-LDOS plots, even if the strong disorder of the $\delta\mu$ term and finite size effects broaden the singularity to some extent. The gap does not close when we vary continuously the disorder strength $w_r$ from zero to its final value and vary the adatom concentration from $n_I=1.0$ to 
its value $n_I=0.2$. This demonstrates that the system is in the same topological phase as the Kane-Mele model\cite{kane1} and that the spectral gap has topological origin.

In conclusion, our approximate analytical and exact numerical calculations based on the graphene/adatom model Eq.\ (\ref{h1}) provide strong evidence for substantial SOC-induced spectral gap opening at the Dirac points in the physically relevant regime of randomly distributed adatoms. Such a gap should be observable in various spectroscopies such as ARPES and STS. In addition, Fourier transform STS should be able to discern unique patterns characteristic of SOC (Fig.\ \ref{fig1}) in the intravalley channel where the signal in pristine graphene is absent due to symmetry considerations.

The authors are indebted to C. Ast, S.A. Burke, A. Damascelli, J.A. Folk, J.E. Hoffman, A. Khademi and B.M. Ludbrook for insightful discussions. This work was supported by NSERC and CIfAR. P. S. thanks the Erasmus Mundus program TEE which made this collaboration possible.


\begin{thebibliography}{10}

\bibitem{moore_rev} J.~E. Moore, \href{http://dx.doi.org/10.1038/nature08916}{Nature (London) {\bf 464}, 194 (2010)}.

\bibitem{hasan_rev} M.Z. Hasan, C.L. Kane, \href{http://dx.doi.org/10.1103/RevModPhys.82.3045}{Rev. Mod. Phys. {\bf 82}, 3045 (2010)}.

\bibitem{qi_rev} X.-L. Qi, S.-C. Zhang, \href{http://dx.doi.org/10.1103/RevModPhys.83.1057}{Rev. Mod. Phys. {\bf 83}, 1057 (2011)}.

\bibitem{franz_book} {\em Topological Insulators,} edited by M. Franz and L. Molenkamp (\href{https://www.elsevier.com/books/topological-insulators/franz/978-0-444-63314-9}{Elsevier,Oxford, England,2013}).

\bibitem{konig} M. Konig, S. Wiedmann, C. Brune, A. Roth, H. Buhmann, L. W. Molenkamp, X.-L. Qi, and S.-C. Zhang, \href{http://dx.doi.org/10.1126/science.1148047}{Science, {\bf 325}, 766 (2007)}.

\bibitem{knez} I. Knez, R.-R. Du, and G. Sullivan, \href{http://dx.doi.org/10.1103/PhysRevLett.107.136603}{Phys. Rev. Lett., {\bf 107}, 136603 (2011)}.

\bibitem{kane1} C. L. Kane and E. J. Mele, \href{http://dx.doi.org/10.1103/PhysRevLett.95.226801}{Phys. Rev. Lett., {\bf 95}, 226801 (2005)}.

\bibitem{weeks1} C. Weeks, J. Hu, J. Alicea, M. Franz, and R. Wu, \href{http://dx.doi.org/10.1103/PhysRevX.1.021001}{Phys. Rev. X, {\bf 1}, 021001 (2011)}.

\bibitem{hu1} J. Hu, J. Alicea, R. Wu, M. Franz, \href{http://dx.doi.org/10.1103/PhysRevLett.109.266801}{Phys. Rev. Lett. {\bf 109}, 266801 (2012)}. 

\bibitem{folk1} A. Khademi and J. Folk (unpublished).

\bibitem{burke1} A. Macdonald and S.A. Burke (unpublished).

\bibitem{dama1}  B.M. Ludbrook and A. Damascelli (unpublished).

\bibitem{shevtsov1} O. Shevtsov, P. Carmier, C. Groth, X. Waintal, D. Carpentier, \href{http://dx.doi.org/10.1103/PhysRevB.85.245441}{\prb {\bf 85}, 245441, (2012)}.

\bibitem{niu1} H. Jiang, Z. Qiao, H. Liu, J. Shi, Q. Niu, \href{http://dx.doi.org/10.1103/PhysRevLett.109.116803}{Phys. Rev. Lett. {\bf 109}, 116803 (2012)}.

\bibitem{shon} N.H.~Shon, and T.~Ando, \href{http://dx.doi.org/10.1143/JPSJ.67.2421}{J. Phys. Soc. Jpn. {\bf 67}, 2421 (1998)}.

\bibitem{doniach1} See e.g.\ {\em Green's Functions for Solid State Physicists,} S. Doniach, E. H. Sondheimer (\href{http://www.worldscientific.com/worldscibooks/10.1142/p067}{Imperial College Press, Reading, MA, 1998}).

\bibitem{crommie1} M.F.~Crommie, C.P.~Lutz, D.M.~Eigler, \href{http://dx.doi.org/10.1038/363524a0}{Nature (London){\bf 363},
524 (1993)}.

\bibitem{davis1} J.~Lee, K.~Fujita, A.R.~Schmidt, C.K.~Kim, H.~Eisaki,
  S.~Uchida, J.C.~Davis, \href{http://dx.doi.org/10.1126/science.1176369}{Science {\bf 325}, 1099 (2009)}, and references therein. 

\bibitem{capriotti1} L.~Capriotti, D.J.~Scalapino and R.D.~Sedgewick, 
\href{http://dx.doi.org/10.1103/PhysRevB.68.014508}{\prb {\bf 68}, 014508 (2003)}.

\bibitem{tami3} T.~Pereg-Barnea and A.H.~MacDonald, \href{http://dx.doi.org/10.1103/PhysRevB.78.014201}{\prb {\bf 78}, 014201 (2008)}.

\bibitem{peskin} See e.g.\ M. E. Peskin and D. V. Schroeder, {\em An 
Introduction to Quantum Field Theory} (Addison-Wesley,Cambridge,MA,1995).

\bibitem{tami1} T.~Pereg-Barnea and M.~Franz, \href{http://dx.doi.org/10.1103/PhysRevB.68.180506}{\prb {\bf 68}, 180506(R) (2003)}; \href{http://dx.doi.org/10.1142/S0217979205026658}{Int.~J.~Mod.~Phys. B {\bf 19}, 731 (2005)}.

\bibitem{bena1} C. Bena, \href{http://dx.doi.org/10.1103/PhysRevLett.100.076601}{\prl {\bf 100}, 076601 (2008)}; \href{http://dx.doi.org/10.1103/PhysRevB.79.125427}{\prb {\bf 79}, 125427 (2009)}.  

\bibitem{mallet1} P.~Mallet, I.~Brihuega, S.~Bose, M.~M.~Ugeda, J.~M.~G\'omez-Rodr\'iguez, K.~Kern, J.~Y.~Veuillen, \href{http://dx.doi.org/10.1103/PhysRevB.86.045444}{\prb {\bf 86}, 045444 (2012)}. 

\end{thebibliography}
\end{document}